\documentclass[a4paper,11pt]{article}
\usepackage{jheppub} % for details on the use of the package, please
\usepackage[T1]{fontenc} % if needed

\title{The effects of deformation parameter on thermal width of moving quarkonia in plasma}

\author[a,1]{J. Sadeghi,}
\author[a,2]{S. Tahery,}
\affiliation[a]{Sciences Faculty, Department of Physics, University of  Mazandaran, Iran}
\emailAdd{j.sadeghi@umz.ac.ir}
\emailAdd{s.tahery@stu.umz.ac.ir}

\abstract{In general we can say that  the thermal width of quarkonia corresponds to imaginary part of it's potential.  Gravity dual of theories give explicit form of potential as $V_{Q\bar{Q}}$. Since there is an explicit formula for $Im V_{Q\bar{Q}}$  one can consider different gravity duals and study the results of  contribution  of various parameters.  Variable gravity duals of moving pair in plasma have different results for potential. Our paper shows that deformation parameter $c$ in warp factor leads to  new results that we present them for arbitrary angles of the pair with respect to  it's velocity.  We compare our results with the case that  no deformation parameter is in metric  background. We will see that the thermal width of the pair increases with increasing deformation parameter. Also, for nonzero values of deformation parameter the pair feels moving plasma in all distances. In addition our results indicate that contribution of deformation parameter leads to larger dissociation length for the moving pair reverse to the effect of the pair's velocity   in the plasma.}

\begin{document} 
\maketitle
\flushbottom
\section{Introduction}
When we want to study $Q \bar{Q}$ interaction we should consider the effect of the medium in  motion of $Q  \bar{Q}$ , because this pair is not produced at rest in QGP. So, the velocity of the pair through the plasma has some effects on its interactions that should be taken into account.
The interaction energy has a finite imaginary part at finite temperature that can be used to estimate the thermal width of the quarkonia \cite{nbma,ybm}. Calculations of Im$V_{Q \bar{Q}}$ relevant to QCD and heavy ion collisions were performed for static $Q \bar{Q}$ pairs using  pQCD \cite{mlop} and lattice QCD  \cite{arth,gaca,gcs} before AdS/CFT.\\
The AdS/CFT  is a correspondence \cite{jmm,ssg,ew,oas} between a string theory in AdS
space and a conformal field theory in physical space-time. It leads
to an analytic semi-classical model for strongly coupled QCD. It has
scale invariance, dimensional counting at short distances and color
confinement at large distances. This theory describes the
phenomenology of hadronic properties and demonstrate their ability
to incorporate such essential properties of QCD as confinement and
chiral symmetry breaking. In the AdS/CFT point of view the $AdS_{5}$
plays important role in describing QCD phenomena. So in order to
describe a confining theory, the conformal invariance of $AdS_{5}$
must be broken somehow. Two strategies AdS/QCD background have been
suggested in the literatures hard-wall model \cite{jee,hrg,hr,eka,jp,ldr} and soft-wall model \cite{ake,sjb,gfde,wdp,hfm,wde,bge,jni,hrgr,hfo,hjk,pcf,ave,aeg,aga,gfd,zab,tbt,avi}. In hard-wall model to impose confinement and discrete
normalizable modes that is to truncate the regime where string modes
can propagate by introducing an IR cutoff in the fifth dimension at
a finite value $z_{0}\sim\frac{1}{\Lambda_{QCD}}$. Thus, the
hard-wall at $z_0$ breaks conformal invariance and allows the
introduction of the QCD scale and a spectrum of particle states,
  they have phenomenological problems, since the obtained
spectra does not have Regge behavior. To remedy this it is necessary
to introduce a soft cut off, using a dilaton field or using a warp
factor in the metric \cite{jee,wde}. These models are called soft wall
models. The soft-wall and hard-wall approach has been successfully
applied to the description of the mass spectrum of mesons and
baryons, the pion leptonic constant, the electromagnetic form
factors of pion and nucleons, etc. On the other hand the study of
the moving heavy quarkonia in space-time with AdS/QCD approach plays
important role in interaction energy \cite{mst,msd,mmk,gac}. By using different
metric backgrounds we see different effects on interaction energy.
\\Evaluation of Im$V_{Q\bar{Q}}$ will yield to determine the suppression of ${Q\bar{Q}}$ in heavy ion collision\cite{sif}.
 The main idea is using boosted frame to have Re $V_{Q\bar{Q}}$ and Im $V_{Q\bar{Q}}$ \cite{fn} for ${Q\bar{Q}}$ in a plasma.\\
From viewpoint of holography, the AdS/CFT correspondence can describe a ``brocken conformal symmetry'',
when one adds a proper deformed warp factor in front of the $AdS_5$ metric structure \cite{jer,gfdt,jba,mkr,tsss,tsa,shm,akek,oan,fzu,gfdet,jpsh,kghm,kghn,ccm,ugek,uek,dfze,hjp,shem,dlis}. So, $e^{cz^2}$ is a positive  quadratic correction with z, the fifth dimension.\\
One natural question is about the connection between the warp factor and the potential $V_{Q\bar{Q}}$. In this work,
the procedure of \cite{sif} is followed to evaluate the imaginary part of potential for an AdS metric background with deformation parameter in warp factor. It is interesting to see `` what
   will happen if meson be in a deformed AdS?''\\
 It is a trend to see the effects of deformation parameter on Re$V_{Q\bar{Q}}$ and Im$V_{Q\bar{Q}}$
  which are evidences for ``usual'' or ``unusual" behavior of meson in compare with the $c=0$ case. 
  As expected in the limit of $c\rightarrow0$, all results are equal to the results of $AdS_5$ case.
 All above informations give us motivation to work on effect of the
  deformation parameter in $AdS$ metric background on real and
  imaginary parts of potential. 
   So, we organized the
  paper as follows. In section 2,  we discuss the case where the pair is moving perpendicularly to the joining axis of the dipole in  deformed AdS,
 we assume this metric background for ${Q\bar{Q}}$ and find some relations
 for real and imaginary parts of potential.
 This  example will be presented with some numerical results for different values of deformation parameter.
 Then we consider general orientation of ${Q\bar{Q}}$ in section 3 and follow the
 procedure as before. Section 4 would be our conclusion and  some suggestions for future work.\\
\section{$ {Q \bar{Q}} $ in an deformed AdS, perpendicular case}
In this section we consider soft-wall metric background with deformation parameter in warp factor at finite temperature case. So, we present  general relations  for real and imaginary parts of potential when the dipole is moving with velocity $\eta$ perpendicularly  to the wind  \cite{sif}.
\\ In our case we apply the general result for deformed AdS, the dual gravity metric will be as:
\begin{equation}
ds^2=e^{2A(z)} [-f(z) dt^2+\Sigma_{i=1}^{i=3} dx^2+\frac{1}{f(z)}
dz^2],
\end{equation}
Where $A (z)=-\ln \frac{z}{R}+\frac{1}{4}cz^2$ and  $f(z)=1-(\frac{z}{z_{h}})^4$. As mentioned before $c$ is deformation parameter and  $R$ is the AdS curvature radius, also $0 \leq z\leq z_{h}$,  $z_{h}=\frac{1}{\pi T}$ and $T$ is boundary field theory's temperature. We have a dynamic dilaton in action for the background and we  write our calculations in string frame. If dilaton is such that it enters directly in the worldsheet action in the form $ \phi R$. May be our concern is about the effect of a nontrivial dilaton profile to the string action. But somewhen people neglect it at the first step \cite{ugkm} and leave it for future study. Then one can check that the integral on the action correspond to worldsheet with higher genus. This means that we are doing string interactions and going to higher order in string perturbation theory. But now, for leading order calculations in genus, we need not to bother with this term even if the geometry has a dynamical dilaton. On the other hand one trace of dynamical dilaton can appear via temperature if we want to calculate it with \cite{GSH} approach. So, the exact temperature will be in hand. But we refer the reader to \cite{dlis} for the reasons that in deformed AdS model with quadratic correction in warp factor the ``temperature'' takes the form of  AdS-SW BH temperature. So, we have a deformed AdS which in the limit $c\rightarrow0$ becomes $AdS_{5}$.  This comparing results help us to underestand the effects of deformation parameter on the physical quantities such as interaction  energy.
  Our  calculations in the
cases of  $LT$, $Re V_{Q\bar{Q}} $ and  $Im V_{Q\bar{Q}}$ give us motivation to compare results between different values of deformation parameter.\\
From metric background (2.1) one can obtain:
\begin{equation}
G_{00}=\frac{R^2}{z^2}[1-(\frac{z}{z_{h}})^4]e^{\frac{cz^2}{2}}
\end{equation}
\begin{equation}
G_{xx}=\frac{R^2}{z^2}e^{\frac{cz^2}{2}}
\end{equation}
\begin{equation}
G_{zz}=\frac{R^2}{z^2}[1-(\frac{z}{z_{h}})^4]^{-1}e^{\frac{cz^2}{2}},
\end{equation}
with  these definitions,
\begin{eqnarray}
\tilde{M}(z)\equiv M(z)\cosh ^2 \eta -N(z)\sinh ^2 \eta\
\end{eqnarray}
\begin{eqnarray}
\tilde{V}(z)\equiv V(z)\cosh ^2 \eta -P(z)\sinh ^2 \eta\
\end{eqnarray}
\begin{equation}
M(z)\equiv G_{00}G_{zz}
\end{equation}
\begin{equation}
V(z)\equiv G_{00}G_{xx}
\end{equation}
\begin{equation}
P(z)\equiv {G_{xx}}^2
\end{equation}
\begin{equation}
N(z)\equiv G_{xx}G_{zz},
\end{equation}
we continue with hamiltonian,
\begin{equation}
H(z)\equiv\sqrt{\frac{\tilde{V}(z)}{\tilde{V}_{\ast}}\frac{\tilde{V}
(z)-\tilde{V}_{\ast}}{\tilde{M}(z)}},
\end{equation}
where $\tilde{V}_{\ast}$ means $\tilde{V}(z_{\ast})$ and $\ast$ is the deepest position
  of the string in the bulk.\\
 The equation of motion and  the boundary conditions of the
string relates $L$(length of the line joining both quarks) with
$z_{\ast}$ as follows,
\begin{equation}
\frac{L}{2}=\int_{r_{\ast}}^{\Lambda}\frac{dr}{H(r)}.
\end{equation}
So, for the corresponding case we have,\\
\begin{equation}
\frac{L}{2}=-\int_{0}^{z_{\ast}}\frac{dz}{H(z)}.
\end{equation}
In order to relation between  $S_{str}$ and $z_{\ast}$  we find the regularized  integral \cite{fn} as,
\begin{eqnarray}
S_{str}^{reg}&=& \frac{T}{\pi \alpha'} \int_{r_*}^{\infty} dr \,\left[\sqrt{\tilde M(r)} \sqrt{\frac{\tilde V(r)}{\tilde V(r_*)}} \left(\frac{\tilde
 V(r)}{\tilde V(r_*)}-1 \right)^{-1/2}-\sqrt{M_0(r)}\right]\nonumber\\ &-&\frac{T}{\pi \alpha'}\int_{r_{h}}^{r_*}dr\,\sqrt{M_0(r)},
\end{eqnarray}
 and we obtain the following results
\begin{equation}
LT=\frac{2}{\pi}y_{h}\sqrt{1-y_{h}^{4}\cosh ^2 \eta} \int_{1}^{\infty} \frac{dy}{\sqrt{(y^4-y_{h}^4)[e^{\frac{cy_{h}^2}{\pi^{2} T^{2}}(\frac{1}{y^2}-1)}(y^4-y_{h}^4\cosh ^2 \eta)-(1-y_{h}^{4}\cosh ^2 \eta)]}}
\end{equation}
where       \quad  \quad $y=\frac{z_{\ast}}{z}$ \quad \quad and\quad  \quad$y_{h}=\frac{z_{\ast}}{z_{h}}$ \\
\begin{eqnarray}
S_{str}^{reg}&=&T^2\frac{\sqrt{\lambda}}{y_{h}} \lbrace\int_{1}^{\infty}dy[\frac{e^{\frac{cy_{h}^2}{\pi^{2} T^{2}}(\frac{1}{y^2}-\frac{1}{2})}(y^4-y_{h}^4\cosh ^2 \eta)}{\sqrt{(y^4-y_{h}^4)[e^{\frac{cy_{h}^2}{\pi^{2} T^{2}}(\frac{1}{y^2}-1)}(y^4-y_{h}^4\cosh ^2 \eta)-(1-y_{h}^{4}\cosh ^2 \eta)]}}\nonumber\\&-&e^{\frac{cy_{h}^2}{\pi^{2} T^{2}y^2}}]-\int_{0}^{1} dy\quad e^{\frac{cy_{h}^2}{\pi^{2} T^{2}y^2}}\rbrace ,
\end{eqnarray}
 Where $\lambda=\frac{R^4}{\alpha^2\prime}$ is  and $ \alpha'$ is the 't Hooft coupling of the gauge theory. Finally,  we find the real part of potential as\\ $Re V_{Q\bar{Q}}=\frac{S_{str}^{reg}}{T}$.\\
 Now we present a derivation of relation for imaginary part of potential from \cite{fn}. The reader can see more details in that reference. From there we can say one should consider the effect of worldsheet fluctuations around the classical configuration $z_c(x)$,
\begin{equation}
z(x) = z_c(x) \rightarrow z(x) = z_c(x) + \delta z (x).
\end{equation}
And then the fluctuations should be taken into account in partition function so one arrives at, 
\begin{equation}
Z_{str} \sim \int \mathcal{D} \delta z(x) e^{i S_{NG} (z_c(x) + \delta z (x))}.
\end{equation}
Then there is an imaginary part of potential in action so , by dividing the interval region of x into $2N$ points where $N\longrightarrow\infty$ that should be taken into account at the end of calculation we arrive at,
\begin{equation}
Z_{str} \sim \lim_{N\to \infty}\int d [\delta z(x_{-N})] \ldots d[ \delta z(x_{N})]  \exp{\left[ i \frac{\mathcal{T} \Delta x}{2 \pi \alpha'} \sum_j \sqrt{M(z_j) (z'_j)^2 + V(z_j)}\right]}.
\end{equation}
Notice that we should expand $z_c(x_j)$ around $x=0$ and keep only terms up to second order of it because thermal fluctuations are important around $z_\ast$ which means $x=0$, 
\begin{equation}
z_c(x_j) \approx z_\ast + \frac{x_j^2}{2} z_c''(0),
\end{equation}
With considering small fluctuations finally we will have,
\begin{equation}
V(z_j) \approx V_* + \delta z V'_* + z_c''(0) V'_* \frac{x_j^2}{2} + \frac{\delta z^2}{2} V''_*,
\end{equation}
where $V_\ast\equiv V(z_\ast)$  and $V'_\ast\equiv V'(z_\ast)$.
With (2.20), (2.21) and (2.19) one can derive (2.22), (2.23) and (2.24),
\begin{equation}
S^{NG}_j = \frac{\mathcal{T} \Delta x}{2 \pi \alpha'} \sqrt{C_1 x_j^2 + C_2}
\end{equation}
\begin{equation}
C_1 = \frac{z_c''(0)}{2} \left[ 2 M_* z_c''(0) + V_*' \right]
\end{equation}
\begin{equation}
C_2 = V_* + \delta z V'_* + \frac{\delta z^2}{2} V''_*.
\end{equation}

 For having $ Im V_{Q\bar{Q}}\neq 0$  the function in the square root of (2.22) should be negative. then, we consider j-th contribution to $Z_{str}$ as,
\begin{equation}
I_j \equiv \int\limits_{\delta z_{j min}}^{\delta z_{j max}} d(\delta z_j) \, \exp{\left[ i \frac{\mathcal{T} \Delta x}{2 \pi \alpha'} \sqrt{C_1 x_j^2 + C_2} \right]},
\end{equation}
For every $\delta z $ between minimum and maximum of it's values which are the roots of  $C_1 x_j^2 + C_2$ in $\delta z $, one leads to $C_1 x_j^2 + C_2 <0$. The extermal value of the function
\begin{equation}
D(\delta z_j) \equiv C_1 x_j^2 + C_2(\delta z_j)
\end{equation}
is,
\begin{equation}
\delta z = - \frac{V'_*}{V''_*}.
\end{equation}
So, $ D(\delta z_j)<0 \longrightarrow  -x_c<x_j<x_c$ leads us to have an imaginary part in square root, where,
\begin{equation}
x_c = \sqrt{\frac{1}{C_1}\left[\frac{V'^2_*}{2V''_*} - V_* \right]}.
\end{equation}
\begin{figure}
\centerline{\includegraphics[width=12cm]{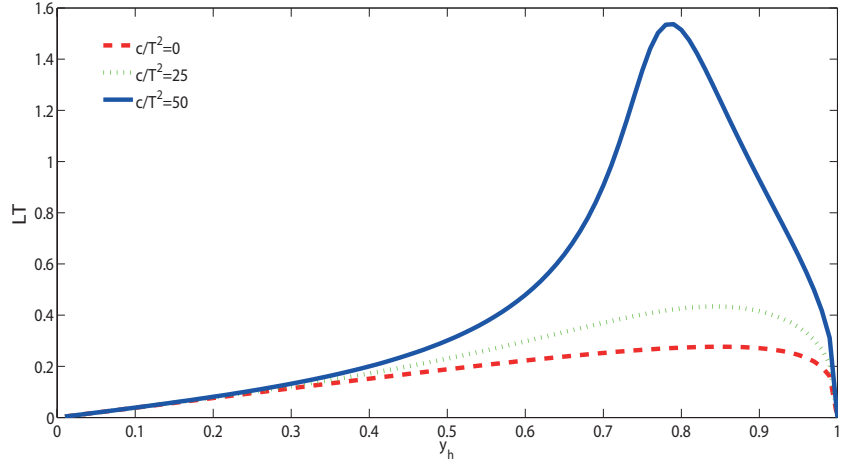}}
\caption{$LT$ as a function of $y_{h}$ at  
 $\eta=0$.  $Q\bar{Q}$  is oriented to the hot wind and
 different values of deformation parameter are contributed. The solid blue curve corresponds to $\frac{c}{T^2}=50$, the dotted green curve to $\frac{c}{T^2}=25$ and the dashed red curve to $\frac{c}{T^2}=0$}
\end{figure}
\begin{figure}
\centerline{\includegraphics[width=12cm]{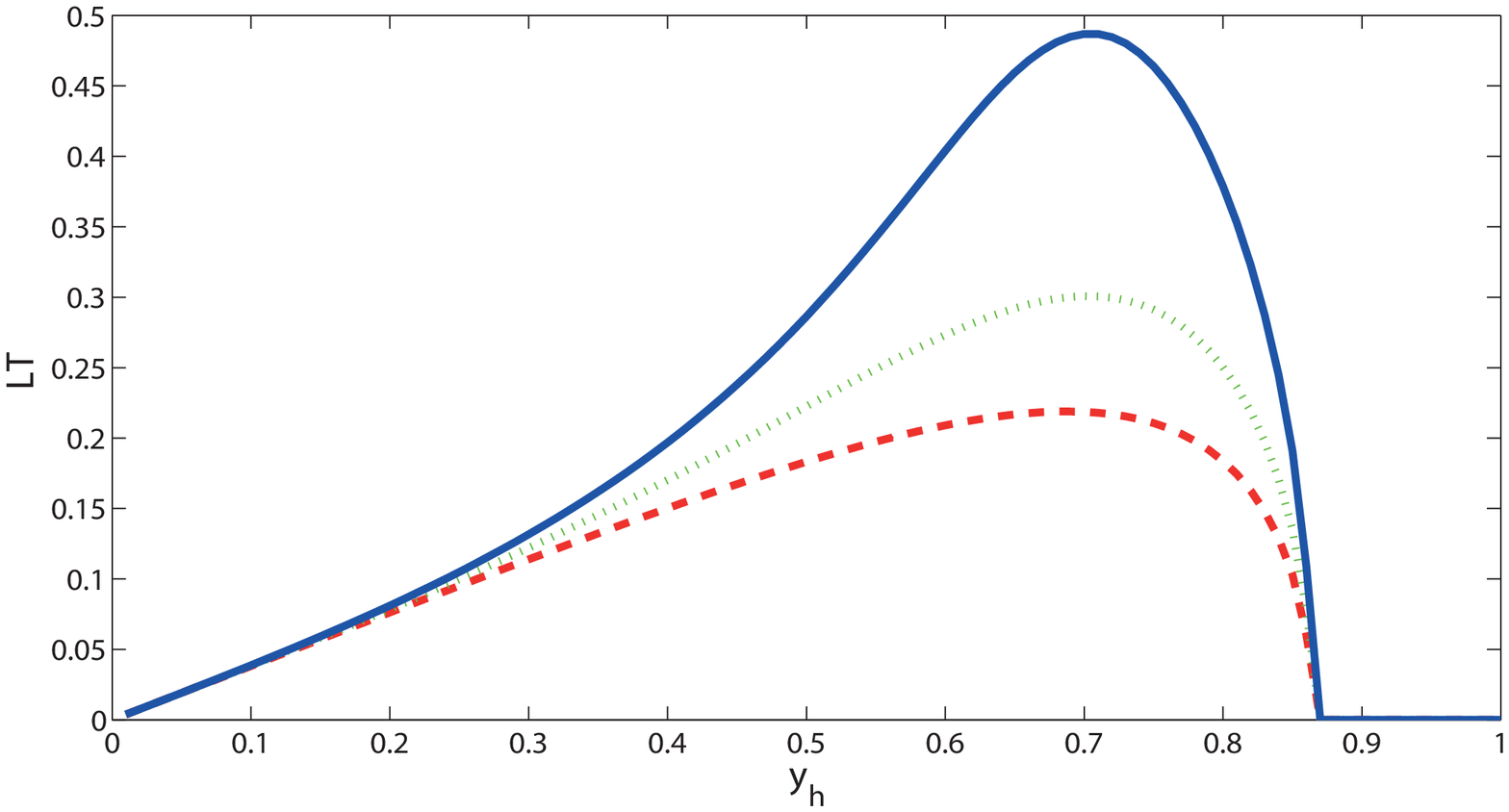}} \caption{$LT$ as a function of $y_{h}$ at  
 $\eta=0.8$.  $Q\bar{Q}$  is oriented to the hot wind and
 different values of deformation parameter are contributed. The solid blue curve corresponds to $\frac{c}{T^2}=50$, the dotted green curve to $\frac{c}{T^2}=25$ and the dashed red curve to $\frac{c}{T^2}=0$}
\end{figure}
\begin{figure}
\centerline{\includegraphics[width=12cm]{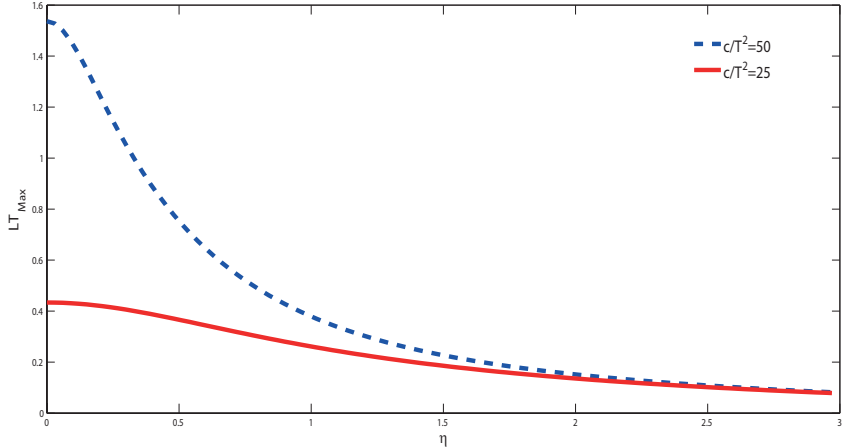}}
\caption{$LT_{max}$ as a function of $\eta$.  $Q\bar{Q}$  is oriented to the hot wind and
 different values of deformation parameter are contributed. The dashed blue curve corresponds to $\frac{c}{T^2}=50$ and the solid red curve to $\frac{c}{T^2}=25$.}
\end{figure}
\begin{figure}
\centerline{\includegraphics[width=12cm]{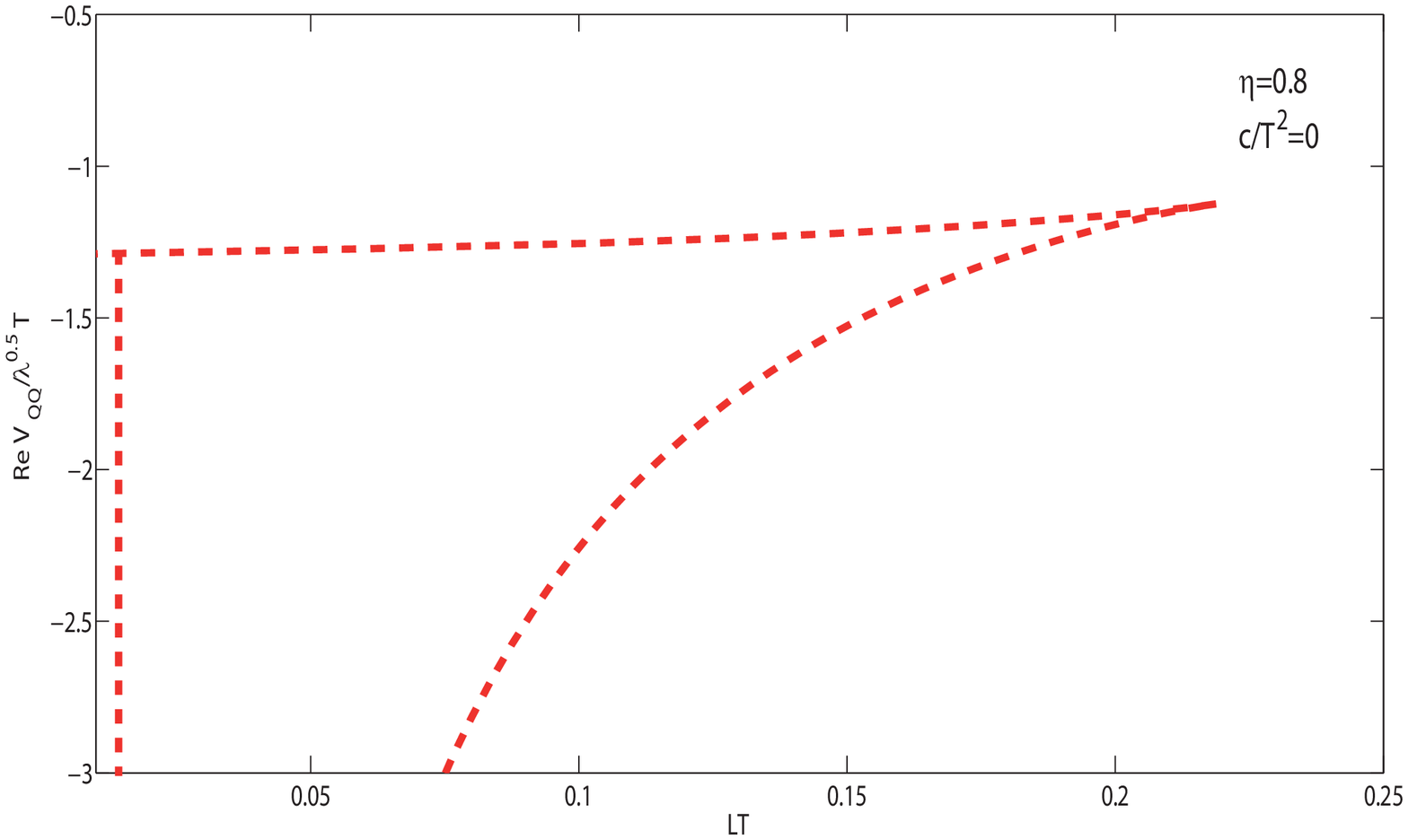}}
\caption{ $ReV_{Q\bar{Q}}$ as a function of $LT$ at a fixed velocity $\eta=0.8$  the pair is oriented to the hot wind and and  deformation parameter is zero}
\end{figure}
\begin{figure}
\centerline{\includegraphics[width=12cm]{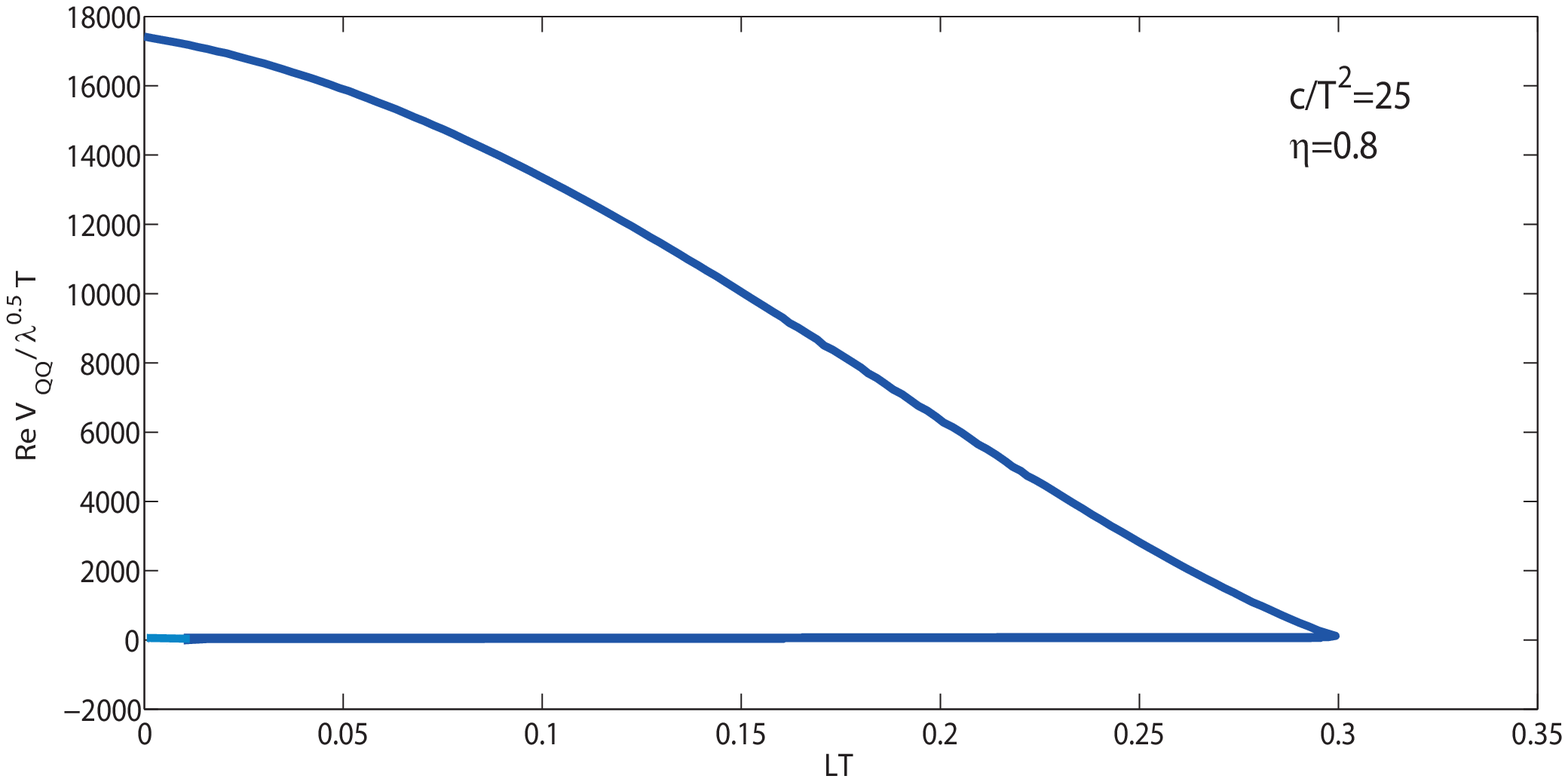}}
\caption{$ReV_{Q\bar{Q}}$ as a function of $LT$  at a fixed velocity $\eta=0.8$  the pair is oriented to the hot wind and scaled deformation parameter is 25}
\end{figure}
\begin{figure}
\centerline{\includegraphics[width=12cm]{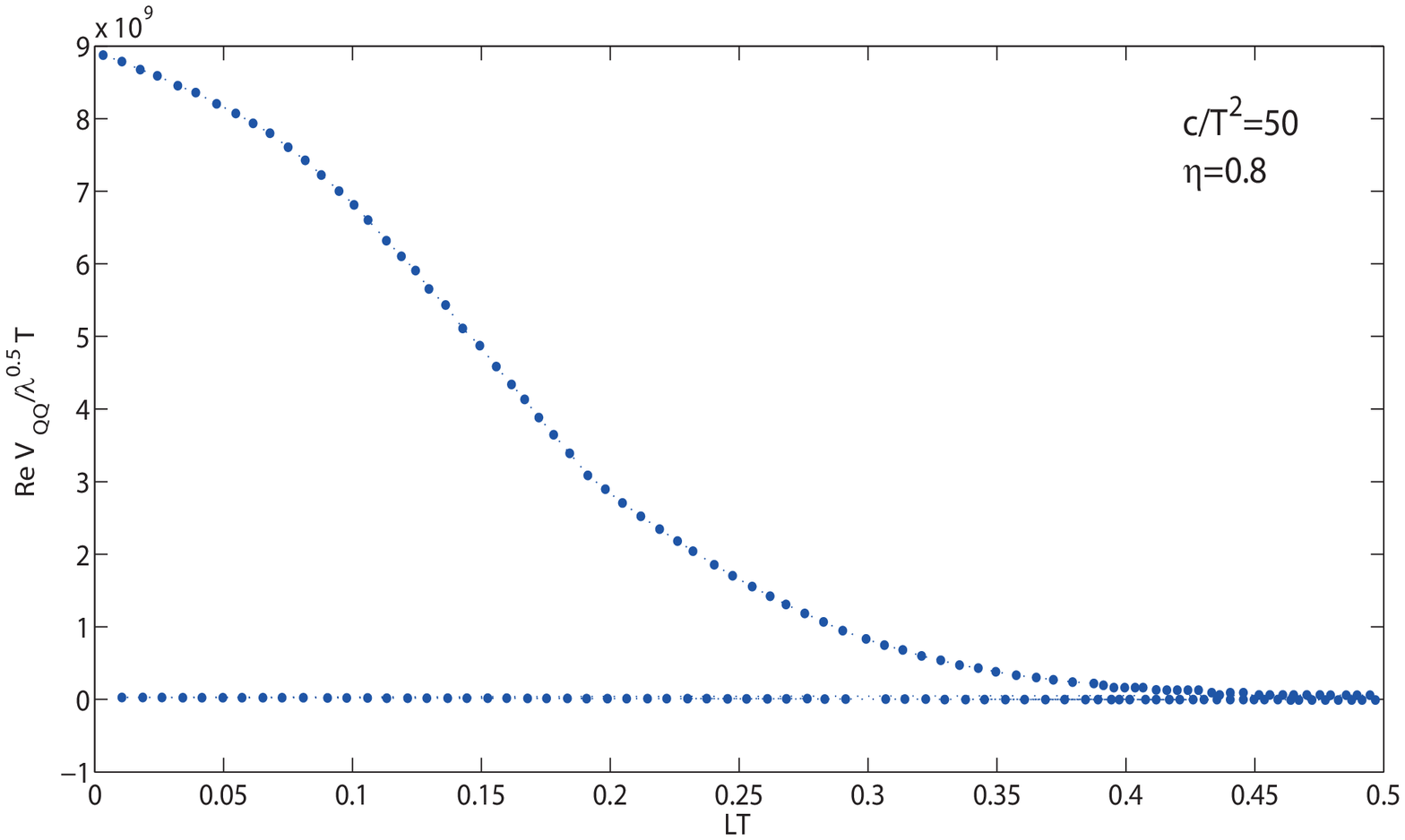}}
\caption{$ReV_{Q\bar{Q}}$ as a function of $LT$  at a fixed velocity $\eta=0.8$  the pair is oriented to the hot wind and scaled deformation parameter is 50}
\end{figure}
\begin{figure}
\centerline{\includegraphics[width=12cm]{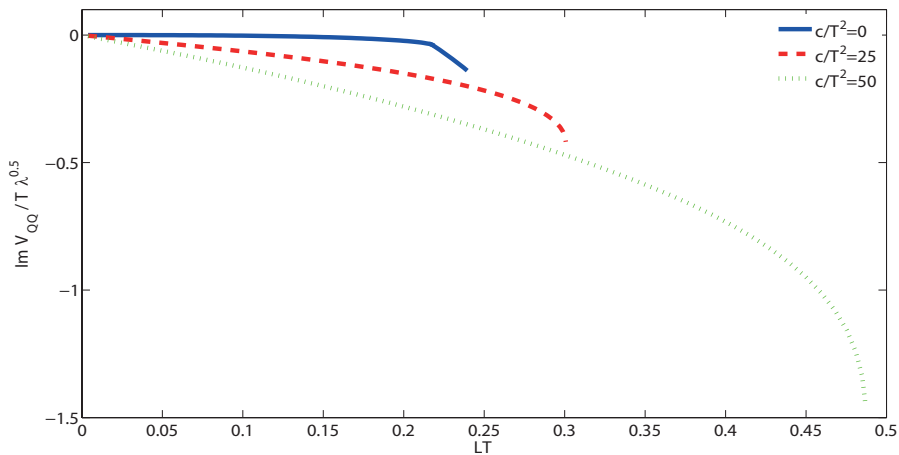}}
\caption{Imaginary part of potential as a function of $LT$, at a fixed velocity $\eta=0.4$.  The pair is oriented to the hot wind and different values of scaled deformation parameter are contributed.  The solid blue curve corresponds to $\frac{c}{T^2}=0$ , the dashed red curve to $\frac{c}{T^2}=25$ and  the dotted green curve to $\frac{c}{T^2}=50$.}
\end{figure}
If the square root in (2.28)  is not real we should take $ x_c=0$. After all these conditions we can approximate $D(\delta z) $ by $ D(-\frac{V'_{\ast}}{V"_{\ast}})$ in $ I_j$,
\begin{equation}
I_j \sim \exp \left[ i \frac{\mathcal{T} \Delta x}{2 \pi \alpha'} \sqrt{C_1 x_j^2 + V_* - \frac{V'^2_*}{2V''_*}} \right].
\end{equation}
The total contribution to the imaginary part, will be in hand with continuum limit. So,
\begin{equation}
\mathrm{Im} \, V_{Q\bar{Q}} = -\frac{1}{2\pi \alpha'} \int\limits_{|x|<x_c} dx \sqrt{-x^2 C_1 - V_* + \frac{V'^2_*}{2V''_*}}\,.
\end{equation}
And finally  after evaluating the integral one can arrive at the expression for imaginary part of potential as,
\begin{equation}
\mathrm{Im} \, V_{Q\bar{Q}} = -\frac{1}{2 \sqrt{2} \alpha'} \sqrt{M_*} \left[\frac{V'_*}{2V''_*}-\frac{V_*}{V'_*} \right].
\end{equation}
Now, we are ready to calculate the imaginary part of potential in case of $\tilde{M}_{\ast} >0$ so according to (2.31)  and with our deformed AdS metric (2.1) we have following relation,
\begin{eqnarray}
\frac{Im V_{Q\bar{Q}}}{\sqrt{\lambda}}&=&-\frac{\pi T}{4\sqrt{2} y_{h}} e^{\frac{cy_{h}^2}{2\pi ^2 T^2}} \sqrt{\frac{1-y_{h}^4\cosh ^2 \eta}{1-y_{h}^4}}\nonumber\\
 &\times &[\frac{\frac{2cy_{h}^2}{\pi ^2 T^2}(1-y_{h}^4\cosh ^2 \eta)+4}{\frac{cy_{h}^2}{\pi ^2 T^2}(2-10y_{h}^4\cosh ^2 \eta) +12}-\frac{(1-y_{h}^4\cosh ^2 \eta)}{2+\frac{cy_{h}^2}{2\pi ^2 T^2}(1-y_{h}^4\cosh ^2
 \eta)}].
\end{eqnarray}
In Fig.1 and 2 we can see the behavior of $ LT$ as a function of $ y_h$ for different values of deformation parameter for this perpendicular case. As we show the maximum of the $LT(y_h)$ which is an indicative of the limit of classical gravity calculations, increases with increasing deformation parameter.\\
 On the contrary, increasing velocity reduces $LT_{max}$ , as it is mentioned in Fig.3. Furthermore, increasing deformation parameter increases $LT_{max}$ which has been used to define a dissociation length for the moving $Q\bar{Q}$ pair.\\
 Fig 4,5 and 6 show the behavior of $ ReV_{Q\bar{Q}} $ as a function of $LT$. As we know for zero value of c in short distances the pair does not feel the moving plasma and upper branch shows saddle point of string action. In compare with that, when c is nonzero pair can feel moving plasma for all values of distances. In addition , we can see with increasing deformation parameter the real part of potential increases and the unphysical curve corresponds to $ q < q_{max}$ and the lower branch is the dominant contribution to the action which corresponds to $ q > q_{max}$.\\
 The imaginary part of the potential corresponds to the dissociation properties of heavy quarkonia. In Fig.7 our results indicate that the thermal width of the pair increases with increasing the deformation parameter at a fixed velocity.

\section{${Q \bar{Q}}$ in an deformed AdS at arbitrary angles}
In this section we extend our calculations for arbitrary angles, it means that orientation of dipole can have any arbitrary angle with respect to velocity vector. As before we extract the real and imaginary parts of potential with method of \cite{sif} , $ \theta$ is the angle of the dipole with respect to the $ X_{d-1}$ and dipole is on the $ (X_1,X_{d-1})$  plane. The boundary conditions are,
\begin{align}
z\left(\pm \frac{L}{2} \sin \theta \right) & = 0 \nonumber \\
X_d \left(\pm \frac{L}{2} \sin \theta \right) & = \pm \frac{L}{2} \cos \theta
\end{align}
And the action is,
\begin{equation}
S_{str} = -\frac{\mathcal{T}}{2\pi \alpha'} \int d \sigma \mathcal{L},
\end{equation}
where the lagrangian is defined as,
\begin{equation}
\mathcal{L} \equiv \sqrt{\left[ M(z) \cosh^2 \eta - N(z)  \sinh^2 \eta \right] z'(\sigma)^2 + V(z) X_d'(\sigma)^2 + \left[ V(z) \cosh^2 \eta - P(z)  \sinh^2 \eta \right] }.
\end{equation}
There are two constants of motion which are,
\begin{equation}
\mathcal{H} \equiv Q \equiv \mathcal{L} - \frac{dz}{d\sigma} \frac{\partial \mathcal{L}}{\partial z'} - \frac{dX_d}{d\sigma} \frac{\partial \mathcal{L}}{\partial X_d'}
\end{equation}
\begin{equation} 
K \equiv \frac{\partial \mathcal{L}}{\partial X_d'},
\end{equation}
with (3.3), (3.4) and (3.5) after some algebra one can arrive at,
\begin{align}
Q^2 \left[ M(z) \cosh^2 \eta - N(z)  \sinh^2 \eta \right] z'(\sigma)^2 + Q^2 V(z) X_{d-1}'(\sigma)^2 + \nonumber \\ + \left[ V(z) \cosh^2 \eta - P(z)  \sinh^2 \eta \right] \left\{Q^2- \left[ V(z) \cosh^2 \eta - P(z)  \sinh^2 \eta \right] \right\} = 0
\end{align}
\begin{align}
K^2 \left[ M(z) \cosh^2 \eta - N(z)  \sinh^2 \eta \right] z'(\sigma)^2 + V(z) (K^2-V(z)) \, X_{d-1}'^2(\sigma) + \nonumber \\ + K^2 \left[ V(z) \cosh^2 \eta - P(z)  \sinh^2 \eta \right]=0\,.
\end{align}
with inserting ${ X'_d}^2$ from (3.6) into (3.7) and doing some manipulations the result is,
\begin{align}
Q^2 V(z) \left[ M(z) \cosh^2 \eta - N(z)  \sinh^2 \eta \right] z'(\sigma)^2 = \nonumber \\ = (V(z)-K^2)  \left[ V(z) \cosh^2 \eta - P(z)  \sinh^2 \eta \right]^2 - V(z) \left[ V(z) \cosh^2 \eta - P(z)  \sinh^2 \eta \right] Q^2 ,
\end{align}
and
\begin{equation}
\label{eq:eqmotionBfin}
Q^2 V^2 (X_{d-1}')^2 = K^2 \left[ V(z) \cosh^2 \eta - P(z)  \sinh^2 \eta \right]^2.
\end{equation}
It is clear that we must have $Z(\sigma=0)=Z_\ast$ , $Z'(\sigma=0)=0$  and $ X_d(\sigma=0)=0$ so,
\begin{equation}
\label{eq:relationUc}
(V_\ast -K^2) (V_\ast \cosh^2 \eta - P_\ast \sinh^2 \eta) - V_\ast Q^2 = 0.
\end{equation} 
 Proceeding by boundary conditions (3.1) and equations of motion (3.8) and (3.9) we arrive at these two relations,
\begin{align}
\frac{L}{2} \sin \theta = &- Q \int_{0}^{z_\ast}\, dz \left\{ \frac{V(z)}{V(z) \cosh^2 \eta - P(z)  \sinh^2 \eta} \right. \times \nonumber \\ & \times \left. \frac{ M(z) \cosh^2 \eta - N(z)  \sinh^2 \eta}{\left[(V(z)-K^2)\left[V(z) \cosh^2 \eta - P(z)  \sinh^2 \eta\right] - V(z) Q^2 \right]} \right\}^{-1/2},
\end{align}
\begin{align}
\frac{L}{2} \cos \theta =- K \int_{0}^{z_\ast}\, dz \, \sqrt{\frac{\left[M(z) \cosh^2 \eta - N(z)  \sinh^2 \eta \right] \left[V(z) \cosh^2 \eta - P(z)  \sinh^2 \eta\right]}{V(z)\left\{(V(z)-K^2)\left[V(z) \cosh^2 \eta - P(z)  \sinh^2 \eta\right] - V(z) Q^2 \right\}}}.
\end{align}
Finally the action is,
\begin{equation}
S = -\frac{\mathcal{T}}{\pi \alpha'} \int_{0}^{z_\ast} \, dz \, \sqrt{\frac{ V(z) \left[M(z) \cosh^2 \eta - N(z)  \sinh^2 \eta \right] \left[V(z) \cosh^2 \eta - P(z)  \sinh^2 \eta\right]}{ \left\{(V(z)-K^2)\left[V(z) \cosh^2 \eta - P(z)  \sinh^2 \eta\right] - V(z) Q^2 \right\}}}\,.
\end{equation}
After regularizing it  we have,
\begin{align}
S_{reg} & =- \frac{\mathcal{T}}{\pi \alpha'} \int_{0}^{z_\ast} \, dz \, \left\{ \sqrt{\frac{ V(z) \left[M(z) \cosh^2 \eta - N(z)  \sinh^2 \eta \right] \left[V(z) \cosh^2 \eta - P(z)  \sinh^2 \eta\right]}{ \left\{(V(z)-K^2)\left[V(z) \cosh^2 \eta - P(z)  \sinh^2 \eta\right] - V(z) Q^2 \right\}}} \right.  \nonumber \\ & \left. - \sqrt{M_0 (z)} \right\} -\frac{\mathcal{T}}{\pi \alpha'} \int_{z_h}^{z_\ast} dz \sqrt{M_0 (z)}.
\end{align}
As before, in the absence of black brane $ M(z)$ for $T=0$ leads to $M_0$ and $ReV_{Q\bar{Q}}=\frac{S_{str}^{reg}}{T}$.
For  imaginary part we have two degrees of freedom $ Z(\sigma)$ and $ X_{d-1}(\sigma)$. The string partition function is,
\begin{equation}
\label{eq:thermalflucpartang}
Z_{str} \sim \int D(\delta z) \, D(\delta X_{d-1}) e^{i S_{str} (\bar{z}+\delta z, \bar{X}_{d-1} + \delta X_d)},
\end{equation}
where fluctuations $\delta z(\sigma)$ and $ \delta X_{d-1}(\sigma)$ are considered with $ \frac{\partial z}{\partial\sigma}\longrightarrow 0$ and $ \frac{\partial X_{d-1}}{\partial\sigma}\longrightarrow 0$.\\
As before with action (3.2) and partitioning  the interval in $ 2N$ subintervals we arrive at,
\begin{equation}
Z_{str} \sim \left( \int_{-\infty}^{\infty} d(\delta z_{-N}) \, d(\delta X_{{d-1},-N}) \right) \cdots \left( \int_{-\infty}^{\infty} d(\delta z_{N}) \, d(\delta X_{{d-1},N}) \right) e^{i \frac{\mathcal{T} \Delta x}{2\pi \alpha'} \mathcal{L}_j},
\end{equation}
and
\begin{equation}
\mathcal{L}_j = \sqrt{\tilde{M}(z(x_j)) (z'(x_j))^2 + V(z(x_j)) (X_{d-1}'(x_j))^2 + \tilde{V} (x_j)}.
\end{equation}
We expand the classical solution $ \bar{z}(0)$ around $\sigma=0$ to quadratic order on $\sigma$. If the string did not sag, then we would have $ X_{d-1}(\sigma)=\frac{\sigma}{\tan \tilde {\theta}}$ and around $\sigma=0$ we will have,
\begin{equation}
X_{d-1}(\sigma) = \frac{\sigma}{\tan \tilde{\theta}} + b \sigma^3 + O (\sigma^5),
\end{equation}
$\tilde{\theta}$ is equal to $\theta$ and b is a constant. Because of the symmetry of the problem under reflections with respect to the origin of the $(X_1,X_d)$ plane, $X_d(\sigma)$ must be an odd function of $ \sigma$ so,
\begin{equation}
X_{d-1}'(\sigma)^2 = \frac{1}{\tan^2 \tilde{\theta}} + \frac{6 b}{\tan \tilde{\theta}} \sigma^2.
\end{equation}
With inserting (3.19) into (3.17), one can arrive at,
\begin{equation}
\mathcal{L}_j = \sqrt{\tilde{C}_1 x_j^2 + \tilde{C}_2},
\end{equation}
with these definition,
\begin{equation}
\tilde{C}_1 \equiv \tilde{M}_\ast\bar{z}''(0)^2+ \frac{1}{2} \left(\frac{V'_\ast}{\tan^2 \tilde{\theta}}+\tilde{V}'_\ast\right)\bar{z}''(0)+\frac{6 b}{\tan \tilde{\theta}} V_\ast
\end{equation}
\begin{equation}
\tilde{C}_2 \equiv  \left(\frac{V_\ast}{\tan^2 \tilde{\theta}}+\tilde{V}_\ast\right) + \left(\frac{V'_\ast}{\tan^2 \tilde{\theta}}+\tilde{V'}_\ast\right) \delta z + \left(\frac{V''_\ast}{\tan^2 \tilde{\theta}}+\tilde{V''}_\ast\right) \frac{(\delta z)^2}{2}.
\end{equation}
As previous section after some algebric calculations the explicite analytical expression for $ImV_{Q\bar{Q}}$ would be,
\begin{equation}
\label{eq:ImFQQang}
\mathrm{Im}\,V_{Q\bar{Q}} = -\frac{1}{4\alpha'}\frac{1}{\sqrt{\tilde{C}_1}} \left[ \frac{\left(\frac{V_\ast'}{\tan^2\tilde{\theta}}+ \tilde{V}_\ast'\right)^2}{2 \left(\frac{V_\ast''}{\tan^2\tilde{\theta}}+ \tilde{V}_\ast''\right)} - \left(\frac{V_\ast}{\tan^2\tilde{\theta}}+ \tilde{V}_\ast\right) \right]\,.
\end{equation}
Again we emphasis that all above derivations about imaginary and real parts of potential are presented in references that we mentioned before, but we follow them here for convenience of the reader. Now we can come back to our main case and follow it with metric (2.1). With using (3.8) and (3.9) we will have,
\begin{equation}
q^2 \left(\frac{dy}{d\tilde{\sigma}}\right)^2 = (y^4 - \cosh^2\eta)((e^{\frac{c}{\pi^{2}T^2y^2}})(y^4-1)-p^2) - q^2(y^4-1) \quad \quad \mathrm{and}
\end{equation}
\begin{equation}
\left(\frac{d\chi}{d\tilde{\sigma}}\right)^2 = \frac{p^2}{q^2} \left(\frac{y^4 - \cosh^2\eta}{y^4-1} \right)^2,
\end{equation}
where we defined the dimensionless variables $y\equiv z_h/z$, $\chi\equiv X_d /z_h$ and $\tilde{\sigma} \equiv \sigma /z_h$ as well as the dimensionless integration constants $q^2 \equiv  Q^2z_h^4/R^4$ and $p^2 =  K^2z_h^4/R^4$
also the boundary conditions become,
\begin{align}
y \left(\pm \pi \frac{LT}{2} \sin \theta \right) & =0 \nonumber \\
\chi \left(\pm \pi \frac{LT}{2} \sin \theta \right) & = \pm \pi \frac{LT}{2} \cos \theta.
\end{align}
So, (3.11), (3.12) and (3.10) lead to (3.27), (3.28) and (3.29),
\begin{equation}
\frac{LT}{2} \pi \sin \theta  =  q \int_{y_\ast}^{\tilde{\Lambda}} \, \frac{dy}{\sqrt{((e^{\frac{c}{\pi^{2}T^2y^2}})(y^4-1)-p^2)(y^4-\cosh^2\eta)-q^2(y^4-1)}} \quad \quad \mathrm{and}
\end{equation}
\begin{equation}
\frac{LT}{2} \pi \cos \theta  =  p \int_{y_\ast}^{\tilde{\Lambda}} \, dy \, \frac{y^4-\cosh^2\eta}{y^4-1} \frac{1}{\sqrt{((e^{\frac{c}{\pi^{2}T^2y^2}})(y^4-1)-p^2)(y^4-\cosh^2\eta)-q^2(y^4-1)}}\,.
\end{equation}
\begin{equation}
((e^{\frac{c}{\pi^{2}T^2y_\ast ^2}})(y_\ast ^4-1)-p^2)(y_\ast ^4-\cosh^2\eta)-q^2(y_\ast ^4-1) = 0\,.
\end{equation}
Therefore the real part of potential is,
\begin{eqnarray}
\frac{\mathrm{Re} \, V_{Q\bar{Q}}}{T\sqrt{\lambda}}& =&  \int_{y_\ast}^{\infty} dy \left[ \frac{e^{\frac{c}{\pi^{2}T^2y^2}}(y^4-\cosh^2\eta)}{\sqrt{(y^4-\cosh^2\eta)(e^{\frac{c}{\pi^{2}T^2y^2}}(y^4-1)-p^2)-q^2(y^4-1)}}-e^{\frac{c}{2\pi^{2}T^2y^2}}\right]\nonumber\\ &-&\int_0^{y_\ast} dy  e^{\frac{c}{2\pi^{2}T^2y^2}}\,.
\end{eqnarray}
And from (3.23) we arrive at imaginary part of potential as,
\begin{eqnarray}
&\frac{\mathrm{Im} \, V_{Q\bar{Q}}}{T\sqrt{\lambda}}&=-\frac{\pi}{4} e^{\frac{c}{2\pi^{2}T^2y_{\ast}^2}}\nonumber\\
&\times&\frac{{\frac{[\frac{2cy_{\ast}}{\pi T}-4\pi Ty_{\ast}^3-\frac{2c}{\pi Ty_{\ast}^3}( \cos^2 \tilde{\theta} + \cosh^2 \eta \sin^2 \tilde{\theta})]^2}{2[20y_{\ast}^2\pi^2 T^2-14c-\frac{4c^2}{\pi ^2T^2y_{\ast}^2}-(\frac{4c^2}{\pi ^2T^2y_{\ast}^6}+\frac{2c}{y_{\ast}^4})( \cos^2 \tilde{\theta} + \cosh^2 \eta \sin^2 \tilde{\theta})]}}-(y_{\ast}^4-( \cos^2 \tilde{\theta} + \cosh^2 \eta \sin^2 \tilde{\theta}))} {\sqrt{y''(0)^2 (\frac{y_{\ast}^4-\cosh^2 \eta}{y_\ast^4-1})+\frac { [\frac{2cy_{\ast}}{\pi^2 T^2} -4y_{\ast} ^3-\frac{2c} {y_{\ast} ^3\pi^2 T^2} ( \cos^2 \tilde{\theta} + \cosh^2 \eta \sin^2 \tilde{\theta})]} {2\sin \tilde{\theta}}y''(0)+ \frac{6 \tilde{b}} {\tan \tilde{\theta}} (y_\ast ^4-1)}}.\nonumber\\
\end{eqnarray}
%\begin{equation}
%\chi(\tilde{\sigma}) = \frac{\tilde{\sigma}}{\tan \tilde{\theta}} + \tilde{b} \tilde{\sigma}^3 + O (\tilde{\sigma}^5)\,.
%\end{equation}
\begin{figure}
\centerline{\includegraphics[width=12cm]{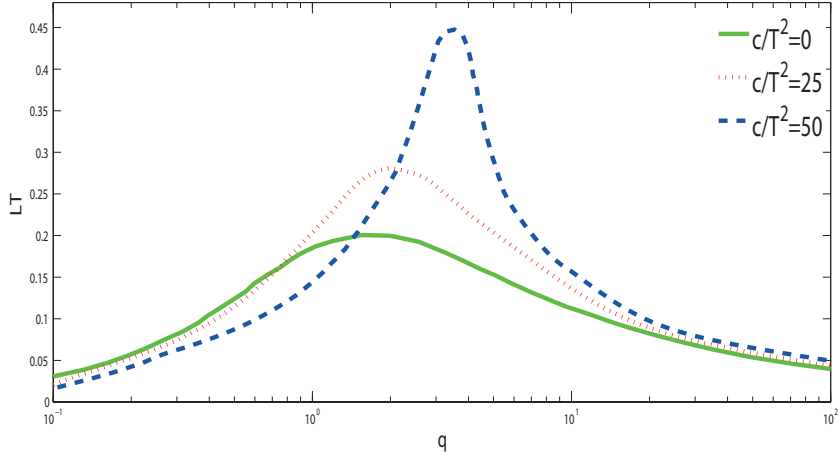}}
\caption{$LT$ as a function of $q$ at  
 $\eta=1$ and $ \theta=\frac{\pi}{3}$. 
 different values of deformation parameter are  contributed. The solid green curve corresponds to $\frac{c}{T^2}=0$, the dotted red curve to $\frac{c}{T^2}=25$ and the dashed blue curve to $\frac{c}{T^2}=50$.}
\end{figure}
\begin{figure}
\centerline{\includegraphics[width=12cm]{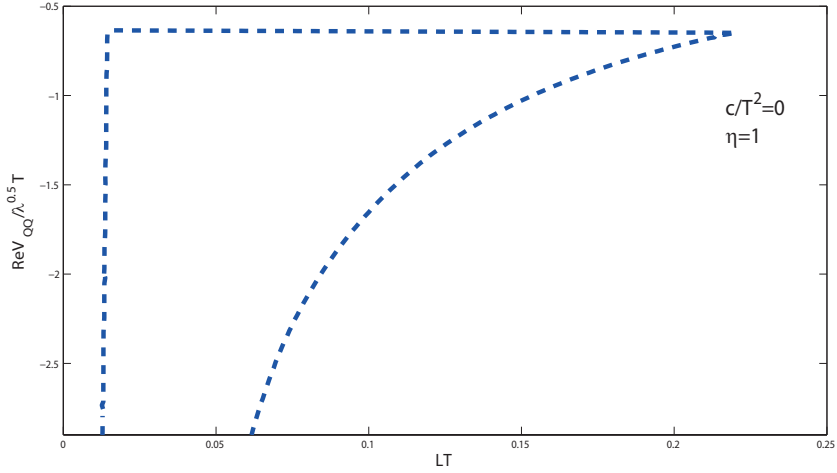}}
\caption{$ReV_{Q\bar{Q}}$ as a function of $LT$ at  
 $\eta=1$ and $ \theta=\frac{\pi}{3}$ and deformation parameter is zero}
\end{figure}
\begin{figure}
\centerline{\includegraphics[width=12cm]{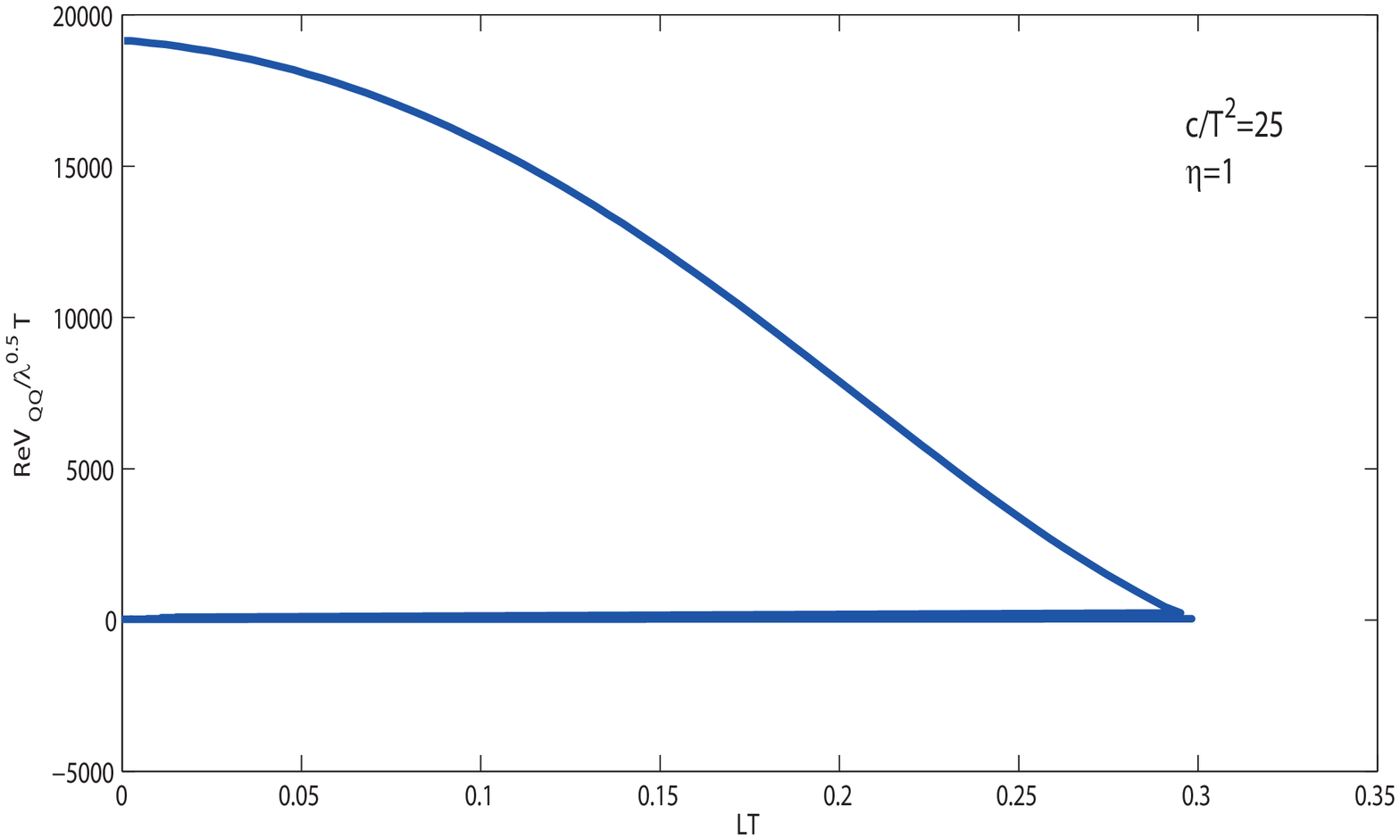}}
\caption{$ReV_{Q\bar{Q}}$ as a function of $LT$ at  
 $\eta=1$ and $ \theta=\frac{\pi}{3}$ and scaled deformation parameter is  $\frac{c}{T^2}=25$}
\end{figure}
\begin{figure}
\centerline{\includegraphics[width=12cm]{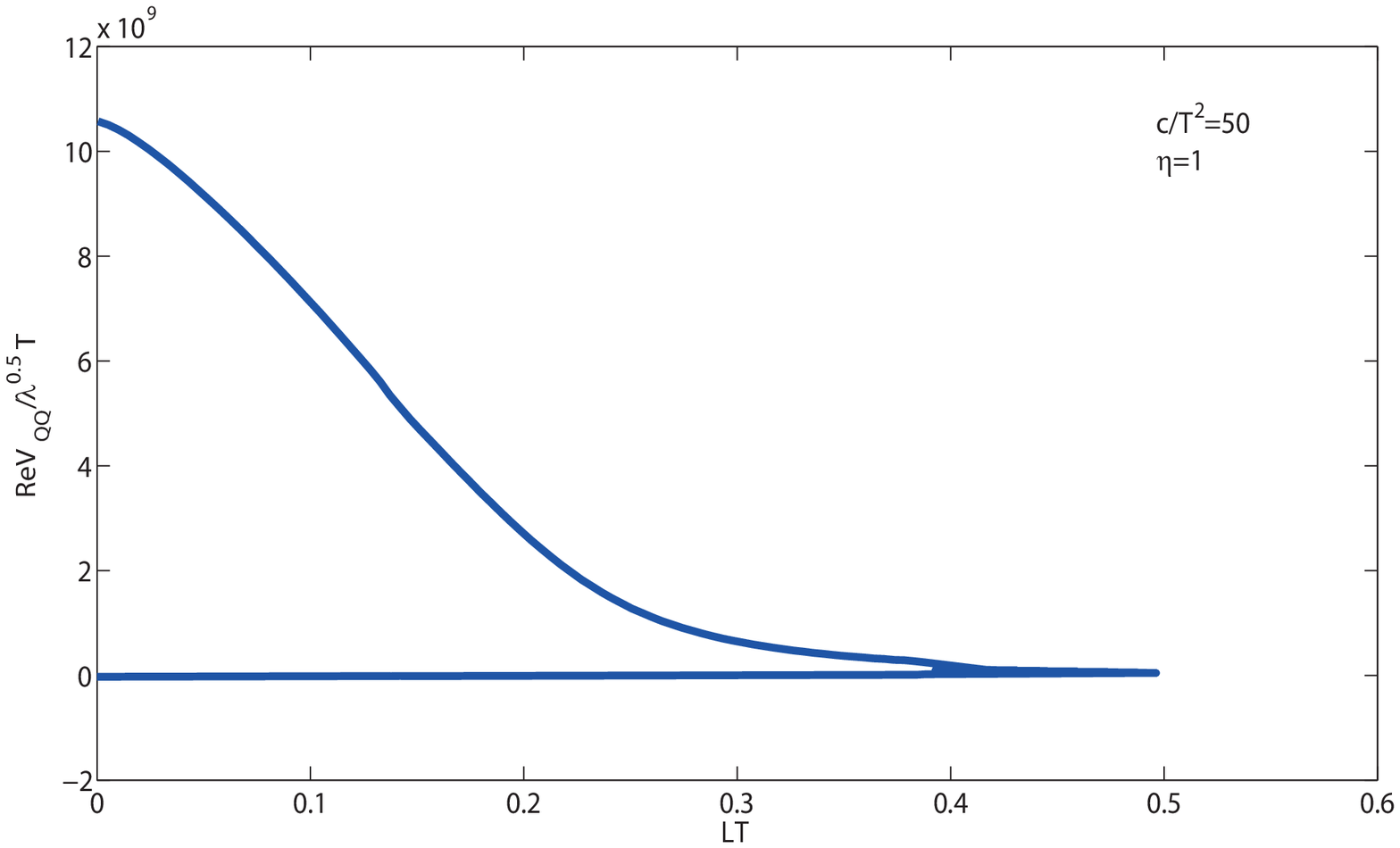}}
\caption{$ReV_{Q\bar{Q}}$ as a function of $LT$ at  
 $\eta=1$ and $ \theta=\frac{\pi}{3}$ and scaled deformation parameter is  $\frac{c}{T^2}=50$}
\end{figure}
\begin{figure}
\centerline{\includegraphics[width=12cm]{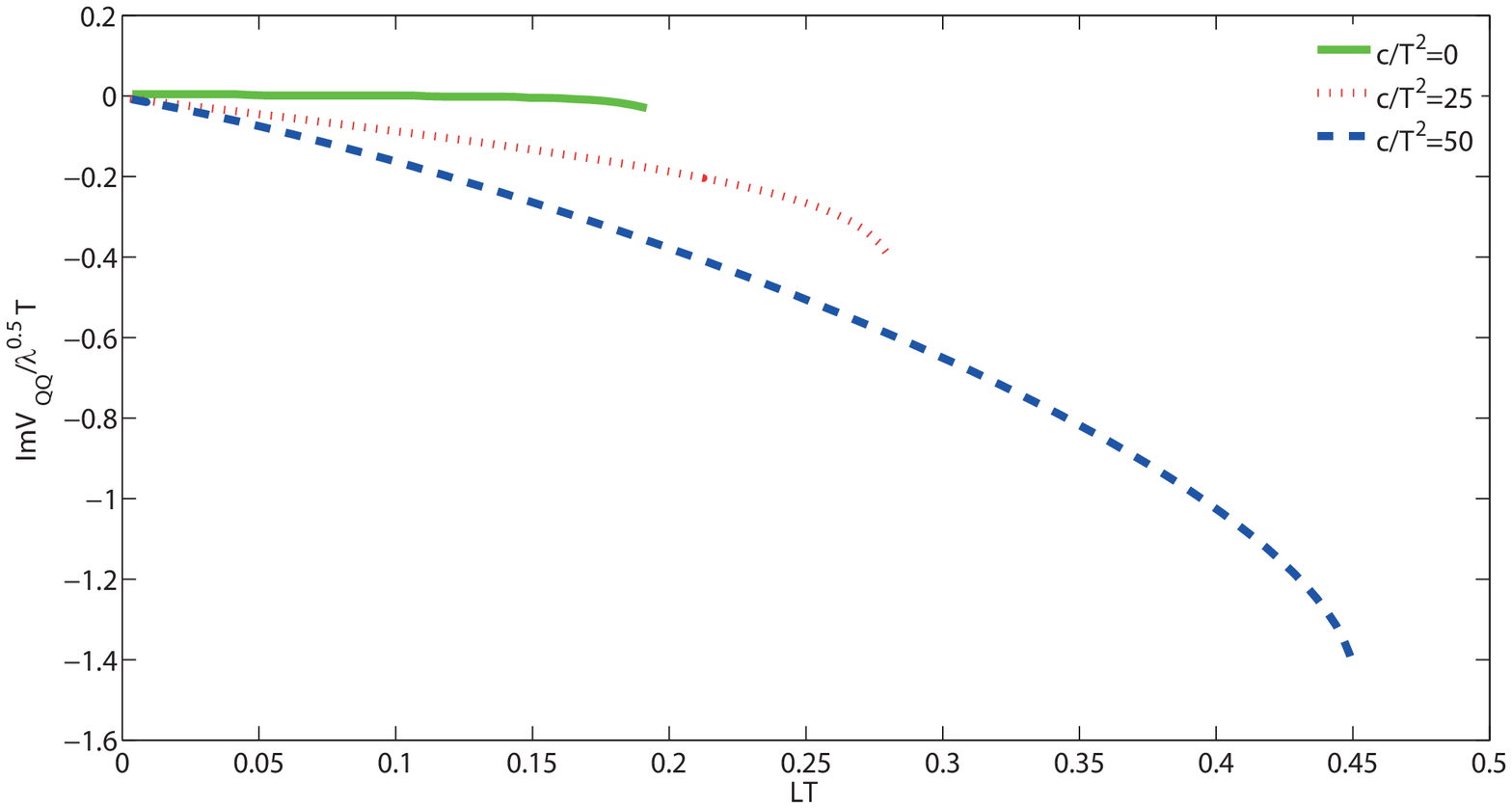}}
\caption{Imaginary part of potential as a function of $LT$, at a fixed velocity $\eta=1$  and $ \theta=\frac{\pi}{3}$.  different values of scaled deformation parameter are contributed.  The solid green curve corresponds to $\frac{c}{T^2}=0$ , the dotted red curve to $\frac{c}{T^2}=25$ and  the dashed blue curve to $\frac{c}{T^2}=50$.}
\end{figure}
We proceed by solving (3.29) numerically to have  $ y_{\ast}$ as a function of q and p, then (3.27), (3.28), (3.30), (3.31) will be functions of p and q. On the other hand for finding p as a function of q,  one can solve (3.27) and (3.28) for fixed $\theta$, after doing all these, $LT$ as a function of q is in hand. Before we start to calculate $Im V_{Q\bar{Q}}$ we should obtain $\tilde{\theta} $, $ y''(0)$ and $ \tilde{b}$. The $\tilde{\theta} $ is obtained from (3.25) at $\tilde{\sigma}=0 $ and $ y=y_{\ast}$. We should solve (3.24) and (3.25) with use of boundary conditions (3.26) , to evaluate $ y''(0)$ and $ \tilde{b}$. After doing all these calculations numerically, with $ y''(0)$ and $ \tilde{b}$ known, we can calculate $Im V_{Q\bar{Q}}$ as a function of q.
So, we will survey $LT(q)$ , also real and imaginary parts of potential as a function of $LT$.\\
The result of cases with a fixed $\eta$ and  different choices of $\theta$ besides a  fixed $\theta$ and  different choices of $\eta$ have been studeid in \cite{sif}, so we proceed by fixing  both of them and choosing different values of deformation parameter.\\
%\begin{eqnarray}
%y_\ast(p,q) &=&\lbrace\frac{1}{ 2e^{\frac{c}{\pi^{2}T^2y_{\ast}^2}}}[( e^{\frac{c}{2\pi^{2}T^2y_{\ast}^2}}+ e^{\frac{c}{2\pi^{2}T^2y_{\ast}^2}}\cosh^2\eta +p^2+q^2)
%+((- e^{\frac{c}{2\pi^{2}T^2y_{\ast}^2}}- e^{\frac{c}{2\pi^{2}T^2y_{\ast}^2}}\cosh^2\eta -p^2-q^2)^2\nonumber\\
%&-&4 e^{\frac{c}{2\pi^{2}T^2y_{\ast}^2}}(p^2 \cosh ^2\eta + e^{\frac{c}{2\pi^{2}T^2y_{\ast}^2}}\cosh ^2\eta %+q^2))^\frac{1}{2}]\rbrace^\frac{1}{4}.
%\end{eqnarray}
In Fig. 8 we show $LT$ as a function of q for a fixed orientation of the dipole , fixed $\eta$ and different values of deformation parameter. we know $LT_{max}$ depends strongly on the rapidity $\eta$ and it decreases with increasing $\eta$ \cite{sif}. In our plots, we can see that  $LT_{max}$ which indicates the limit of validity of classical gravity calculation, increases with increasing deformation parameter.\\
In Figs. 9,10 and 11 we present $ ReV_{Q\bar{Q}}$ as a function of LT. We can see for small values of LT which means short distances or small temperatures, there is a difference between $c=0$ and $c\neq 0$ cases. As we expected, when deformation parameter contributes to the calculation, the interaction of the pair is relevant with plasma, it is similar to the result of perpendicular case. The other point is that real part of potential  has no intense alteration with varying angle for any value of deformation parameter.\\
In Fig. 12 we can see $ImV_{Q\bar{Q}}$ as a function of LT. It shows that for angle $\theta < \frac{\pi}{2} $  with decreasing angle, the imaginary part of potential becomes smaller for any values of deformation parameter.
\section{Concolusion}
In this article, we have used the method of \cite{sif} to investigate the real and imaginary parts of potential for moving heavy quarkonia in plasma with a gravity dual which has deformation parameter in warp factor. At the first step we considered  $Q\bar{Q}$ pair oriented perpendicularly to the hot wind and after that we extended all calculations to arbitrary angles. We saw that for both perpendicular and arbitrary angle cases, the limit of classical gravity calculation increases with increasing deformation parameter. Also for nonzero values of c the pair feels moving plasma even in short distances, but for $c=0$ case the pair does not feel moving plasma at some small values of LT as we expected. We indicated when nonzero values of deformation parameter contribute to the imaginary part of  potential, the thermal width of quarkonia increases with increasing deformation parameter. Results of perpendicular case in compare with arbitrary angle $\theta < \frac{\pi}{2} $ showed that with decreasing angle, the imaginary part of potential becomes smaller for any values of deformation parameter, but real part of potential  has no intense alteration with varying angle for any value of deformation parameter. \\
 Another interesting problem is instead of using
 the soft wall model we use  hyperscaling violation metric background  and
 discuss the moving mesons and investigate real and imaginary parts of
 potential. This problem with corresponding metric background for
 the moving meson  in plasma media is in hand.
\newpage 
\textbf{Acknowledgement}\\
The authors are grateful very much to S. M. Rezaei for support and valuable activity in numerical calculations.

\end{document}